\documentstyle[preprint,eqsecnum,aps]{revtex}
\renewcommand{\baselinestretch}{2}
\begin{document}
\tighten
\draft
\title{Quark confinement and color transparency in a gauge-invariant formulation of QCD}
\author{Lusheng Chen\thanks{chen@phys.uconn.edu} and Kurt Haller\thanks{khaller@uconnvm.uconn.edu}}
\address{Department of Physics, University of Connecticut, Storrs, Connecticut
06269}

\maketitle
\begin{abstract}
We examine a nonlocal interaction that results from expressing the QCD Hamiltonian entirely in 
terms of gauge-invariant quark and gluon 
fields. The interaction couples one quark color-charge density to another, much as electric charge 
densities are coupled to each other by the Coulomb interaction in QED. In QCD, this 
nonlocal interaction also couples quark 
color-charge densities to gluonic color. We show how the leading part of the 
interaction between quark color-charge densities vanishes when the participating quarks 
are in a color singlet configuration, and that, for singlet configurations, the residual interaction weakens 
as the size of a packet of quarks shrinks. Because of this effect, color-singlet packets of quarks
should experience final state interactions that increase in strength as these packets expand in size. 
For the case of an SU(2) model of QCD based on the {\em ansatz} that the gauge-invariant gauge field is 
a hedgehog configuration, we show how the infinite series that 
represents the nonlocal interaction between quark color-charge 
densities can be evaluated nonperturbatively, without 
expanding it term-by-term. 
We discuss the implications of this model for QCD with SU(3) color and a gauge-invariant gauge field 
determined by QCD dynamics.
\end{abstract}
\bigskip \bigskip \bigskip

\newpage

\narrowtext

\renewcommand{\baselinestretch}{2}
\section{Introduction}
\label{Intro}
In previous work,\cite{BCH1,CBH2,BCH3} we developed a gauge-invariant formulation of QCD 
in which the QCD Hamiltonian is 
described as a functional of gauge-invariant operator-valued quantities --- spinor (quark) fields, gauge fields 
and their canonical momenta. One outgrowth of this work was the demonstration of an important 
similarity between QED and QCD. In a formulation of QED in which the charged spinor field 
represents a gauge-invariant quantity, 
 a nonlocal interaction between charge densities --- the Coulomb interaction --- 
appears as part of the Hamiltonian. On the other hand, as is well-known, there is no explicit nonlocal interaction in 
standard versions of QED in covariant gauges in which the charged spinor, $\psi$, represents a 
gauge-dependent field. However, when covariant-gauge QED is transformed to a representation in which  Gauss' law 
is implemented and $\psi$ becomes a gauge-invariant charged field, the same nonlocal 
Coulomb interaction that is seen in the Coulomb gauge appears explicitly in that case as well,
even though the covariant-gauge condition continues to apply to it.\cite{khelqed} 
These observations illustrate that, when only gauge-invariant fields are used in constructing the Hamiltonian ---
in a variety of gauges, including but not limited to the Coulomb gauge ---
the interactions between charged fields 
and pure-gauge parts of gauge fields vanish, and a nonlocal interaction 
between charge densities appears in their stead. The fact that the Coulomb interaction is, by far, the most 
important electrodynamic force in the low-energy regime, provides strong incentive for formulating QCD in 
terms of gauge-invariant fields, to explore the implications of these observations for 
our understanding of QCD. \bigskip

The organization of this paper is as follows: In the section following this Introduction, we review and expand on 
relevant material from our previous work,
which provides a foundation for describing QCD in terms of gauge invariant fields. This part of our discussion is exact, 
and requires no approximations. In a later section, we will use these results to suggest how quark confinement and color 
transparency could be understood as a consequence of the gauge-invariant formulation of the QCD Hamiltonian. And 
finally, we will invoke an {\em ansatz} about the form of the gauge-invariant gauge field to illustrate a 
technique for evaluating the nonlocal interaction nonperturbatively.\bigskip
  
\section{The QCD Hamiltonian as a functional of gauge-invariant fields}
\label{sec:QCDHam}
We will review here some technical developments that we discussed previously,\cite{BCH1,CBH2,BCH3} 
and that are essential for the investigation reported in this work. One of these 
developments is the construction of a set of gauge-invariant quark and gluon operator-valued fields. 
Another is the 
transformation of the QCD Hamiltonian to a representation in which it is 
expressed in terms of these gauge-invariant 
fields. In our work, the QCD Hamiltonian, $\tilde{H}$, is expressed in terms of 
gauge-invariant operator-valued fields
in a representation in which $\psi$ designates the gauge-invariant quark field. It is given by

\begin{equation}
\tilde{H}=\int d{\bf r}\left[ \ {\textstyle \frac{1}{2}}
\Pi^{a}_{i}({\bf r})\Pi^{a}_{i}({\bf r})
+  {\textstyle \frac{1}{4}} F_{ij}^{a}({\bf r}) F_{ij}^{a}({\bf r})+
{\psi^\dagger}({\bf r})\left(\beta m-i\alpha_{i}
\partial_{i}\right)\psi({\bf r})\right] + \tilde{H}^{\prime}\,.
\label{eq:HQCDN}
\end{equation}
$\tilde{H}^{\prime}$ describes interactions involving the gauge-invariant quark field, and can be expressed as 
\begin{equation}
\tilde{H}^{\prime}=\tilde{H}_{j-A}+\tilde{H}_{\cal G}+\tilde{H}_{LR}\,.
\end{equation}
As shown in Ref.\cite{BCH3}, ${\tilde{H}}_{\cal G}$ vanishes in the representation (the so-called ${\cal N}$ 
representation) in which this Hamiltonian is described, and will not be given any further consideration in this work. 
\label{eq:Hprime} 
$\tilde{H}_{j-A}$ describes the interaction of the gauge-invariant gauge field with the 
transverse gauge-invariant quark color-current density, and is given by 
\begin{equation}
\tilde{H}_{j-A}=-\,g\int d{\bf r}\,{\psi^\dagger}({\bf r})\alpha_{i}{\textstyle\frac{\lambda^h}{2}}\psi({\bf r})\,
A_{{\sf GI}\,i}^{h}({\bf r})\,;
\label{eq:HJA}
\end{equation}
${\tilde{H}}_{LR}$ is the nonlocal interaction 
\begin{equation}
{\tilde{H}}_{LR}=H_{g-Q}+H_{Q-Q}
\label{eq:HLR}
\end{equation} 
 with\footnote{Eq.~(\ref{eq:HQCDglue}) corrects a typographical error in Eq. (23) in Ref.\cite{BCH3}.}
\begin{eqnarray}      
H_{g-Q}&=&\int d{\bf r}\left\{\,+{\sf Tr}\left[\,\sum_{r=0}
g^{r+1}(-1)^{r}f^{\vec{\delta}d h}_{(r)}f^{d\sigma e}
\Pi^{a}_{i}({\bf r})\,{\textstyle\frac{\lambda^a}{2}}A_{{\sf GI}\,i}^{\sigma}({\bf r})\,
V_{\cal{C}}^{-1}({\bf{r}})
{\textstyle\frac{\lambda^e}{2}}V_{\cal{C}}({\bf{r}})
{\textstyle\frac{1}{\partial^{2}}}
\left({\cal T}_{(r)}^{\vec{\delta}}({\bf r})
j_0^{h}({\bf r})\right)\,\right]\right.\;
\nonumber \\
&&\;\;\;\;\;\;+ {\sf Tr}\left.\left[\,\sum_{r=0}
g^{r+1}(-1)^{r}f^{\vec{\delta}d h}_{(r)}f^{d\sigma e}
{\textstyle\frac{1}{\partial^{2}}}
\left({\cal T}_{(r)}^{\vec{\delta}}({\bf r})
j_0^{h}({\bf r})\right)V_{\cal{C}}^{-1}({\bf{r}})
{\textstyle\frac{\lambda^e}{2}}V_{\cal{C}}({\bf{r}})
A_{{\sf GI}\,i}^{\sigma}({\bf r})
\Pi^{b}_{i}({\bf r}){\textstyle\frac{\lambda^b}{2}}\,\right]\right\}
\label{eq:HQCDglue}
\end{eqnarray} and
\begin{equation}
H_{Q-Q}= \left\{{\textstyle \frac{1}{2}}\sum_{r=0}
\sum_{r^\prime =0}
g^{r+r^\prime}(-1)^{r+r^\prime}
f^{\vec{\delta}d h}_{(r)}f^{\vec{\delta}^\prime d h^\prime}_{(r^\prime)}
{\textstyle\frac{\partial_{i}}{\partial^{2}}}
\left({\cal T}_{(r)}^{\vec{\delta}}({\bf r})
j_0^{h}({\bf r})\right)
{\textstyle\frac{\partial_{i}}{\partial^{2}}}
\left({\cal T}_{(r^\prime)}^{\vec{\delta}^\prime}({\bf r})
j_0^{h^\prime}({\bf r})\right)\,\right\}\,;
\label{eq:HQCDNVCRL}
\end{equation}
here $A_{{\sf GI}\,i}^{\sigma}({\bf r})$ is the transverse, gauge-invariant gauge field constructed in 
Ref.\cite{CBH2} and $j_0^{h}$ is the gauge-invariant 
quark color-charge density $j_0^{h}=g{\psi}^{\dagger}({\lambda}^h/2){\psi}\,$; 
$f^{\vec{\alpha}\beta\gamma}_{(\eta)}$ is
the chain of structure constants
\begin{equation}
f^{\vec{\alpha}\beta\gamma}_{(\eta)}=f^{\alpha[1]\beta b[1]}\,
\,f^{b[1]\alpha[2]b[2]}\,f^{b[2]\alpha[3]b[3]}\,\cdots\,
\,f^{b[\eta-2]\alpha[\eta-1]b[\eta-1]}f^{b[\eta-1]\alpha[\eta]\gamma}
\label{eq:fproductN} \end{equation}
summed over repeated Lie group indices; 
$V_{\cal{C}}^{-1}({\bf{r}}){\textstyle\frac{\lambda^e}{2}}V_{\cal{C}}({\bf{r}})$ 
is a quantity that 
transforms like an SU(N) vector (where $N=3$ for QCD and $N=2$ for Yang-Mills theory); 
and 
$\left({\cal T}_{(r)}^{\vec{\delta}}({\bf r})
j_0^{h}({\bf r})\right)$ is given by 
\begin{equation}
{\cal T}_{(r)}^{\vec{\delta}}({\bf r})j_0^{a}({\bf{r}})=
A_{{\sf GI}\,j(1)}^{{\delta}(1)}({\bf r})\,
{\textstyle\frac{\partial_{j(1)}}{\partial^{2}}}
\left(A_{{\sf GI}\,j(2)}^{{\delta}(2)}({\bf r})\,
{\textstyle\frac{\partial_{j(2)}}{\partial^{2}}}
\left(\cdots\left(A_{{\sf GI}\,j(r)}^{{\delta}(r)}({\bf r})\,
{\textstyle\frac{\partial_{j(r)}}{\partial^{2}}}
\left(j_0^{a}({\bf{r}})\right) \right)\right)\right).
\label{eq:calT}
\end{equation}
Eq.~(\ref{eq:HQCDNVCRL}) can be understood as the non-Abelian analog of the Coulomb interaction
in QED, and to have a structure similar to that of the Coulomb interaction, when the quantity 
${\cal K}_0^d({\bf r})=\sum_{r=0}f^{\vec{\delta}d h}_{(r)}(-g)^{r}\left({\cal T}_{(r)}^{\vec{\delta}}
({\bf r})j_0^{h}({\bf r})\right)$ is substituted for the Abelian $j_0({\bf r})$ that represents 
the electric charge density in QED. The same ${\cal K}_0^d({\bf r})$ also participates in the 
nonlocal interaction described in Eq.~(\ref{eq:HQCDglue}), where it 
couples to ``glue''-color, ${\sf K}_g^d$ given by
\begin{equation} 
{\sf K}_g^d({\bf r})=gf^{d\sigma e}\,{\sf Tr}\left[V_{\cal{C}}^{-1}({\bf{r}})
{\textstyle\frac{\lambda^e}{2}}V_{\cal{C}}({\bf{r}}){\textstyle\frac{\lambda^b}{2}}\right]
A_{{\sf GI}\,i}^{\sigma}({\bf r})
\Pi^{b}_{i}({\bf r})\,.
\label{eq:Kglue}
\end{equation}

Before proceeding with an application of our earlier work to a discussion of quark color confinement and color 
transparency, we need to clarify some questions about the formalism we have constructed. Our first remark addresses the 
fact that gauge-invariant fields are not generally unique, either in Abelian theories, such as QED, or in QCD. 
Since the Gauss' law operator is the generator of infinitesimal gauge transformations, any unitary operator that 
commutes with the Gauss' law operator can be applied to a gauge-invariant state, or can be used to unitarily 
transform gauge-invariant 
operators, without interfering with their gauge invariance. A number of such unitary operators that commute with the 
Gauss' law operator for QED were displayed in Ref.\cite{khelqed}, and further operators, that 
transform gauge-invariant operators 
in other Abelian gauge theories without interfering with their gauge invariance,  
have also been constructed.\cite{khelmcs} 
Constructing such operators for QCD would be more difficult, but there is 
little doubt that it would be possible to do so. We 
therefore need to address the non-uniqueness of the gauge-invariant operators used in our formulation. \bigskip

In previous work,\cite{khelqed,khelmcs}, we have used the following criterion for accepting a gauge-invariant field as
useful for a gauge-invariant formulation of a gauge theory; 
we propose to apply that same criterion in this case as well. We require that when the interactions 
between the pure gauge degrees of freedom and gauge-invariant matter fields have been eliminated in favor of 
nonlocal interactions involving gauge-invariant matter fields --- $H_{Q-Q}$ and $H_{g-Q}$ in the present case 
--- the remaining interactions 
are restricted to interactions of the gauge-invariant gauge field with transverse current densities only. 
The unavailability of a longitudinal component of the gauge-invariant gauge field in $H_{j-A}$ precludes the formation, 
through the operation of current conservation, of a 
coupling between charge density and a longitudinal gauge field. 
No further interactions between matter-field charge densities can therefore be 
transmitted through virtual loops of gauge-invariant gauge field components, making the nonlocal interactions
the dominant features in a description of low-energy dynamics. \bigskip

In most gauge-invariant formulations of QED, 
such as those in the Coulomb gauge, the Coulomb interaction, $-\,\frac{1}{2}{\int}d{\bf r}
j_0({\bf r}){\nabla}^{-2}j_0({\bf r})$, and the interaction of
the transverse (gauge-invariant) part of the gauge field with the current density, constitute  
the interaction Hamiltonian. That formulation satisfies the criterion we have proposed. In axial ($A_3=0$) gauge QED,
however, the nonlocal interaction is given by $-\,\frac{1}{2}{\int}d{\bf r}j_0({\bf r}){\partial}^{-2}_{3}j_0({\bf r})$
and {\em is} accompanied by further interactions of the gauge field with charge as well as current densities. 
Further unitary transformations are then necessary to eliminate contributions from these additional interactions, 
through virtual photon loops, to the formation of forces between static charges. Appropriately chosen 
unitary transformations not only eliminate these 
interactions among static charges through virtual loops, but also transform 
$-\,\frac{1}{2}{\int}d{\bf r}j_0({\bf r}){\partial}^{-2}_{3}j_0({\bf r})$ to the Coulomb interaction 
$-\,\frac{1}{2}{\int}d{\bf r}j_0({\bf r}){\nabla}^{-2}j_0({\bf r})$ and thereby restore rotational 
symmetry to the nonlocal interaction, and 
still maintain the axial gauge condition.\cite{khelqed} It should also be noted that, in QED, our proposed criterion 
requires that gauge-invariance be imposed on matter fields through unitary transformations that do not 
dress charged fields with transverse propagating photons that have no role in implementing Gauss' law.\cite{khelqed} 
A similar requirement, that charged particles not be dressed spuriously with transverse, propagating photons,  
was proposed by Haagensen and Johnson as necessary for avoiding ``false confinement'' in Wilson loop 
calculation for QED.\cite{johnson4}\bigskip

When applied to QCD, we note that the nonlocal interactions in ${\tilde H}^{\prime}$ are $H_{Q-Q}$ and $H_{g-Q}$, 
and that ${\tilde H}_{j-A}$ describes the interaction of  
the transverse, gauge-invariant gluon field $A_{{\sf GI}\,i}^{h}({\bf r})$ 
with quark color {\em transverse current} densities only. 
${\tilde H}_{\cal G}$ is a term in the transformed Hamiltonian in which ${\cal G}^a$ appears either on the extreme 
left or right. ${\cal G}^a$ represents 
the Gauss's law operator in the representation in which the transformed Hamiltonian ${\tilde H}$
is expressed. ${\cal G}^a$ always annihilates either the ``bra'' or ``ket'' state vector of matrix elements taken 
between allowed states. And, since ${\cal G}^a$ commutes with ${\tilde H}$, state vectors that implement Gauss's law 
initially, will continue to do so as they time-evolve under the influence of ${\tilde H}$.
The nonlocal  $H_{Q-Q}$ and $H_{g-Q}$ therefore describe {\em all} the interactions between static quarks. 
No interactions between quark color charge densities can be 
transmitted through virtual loops 
generated by the gauge-invariant gluon field $A_{{\sf GI}\,i}^{h}({\bf r})$ without involving $H_{Q-Q}$ or $H_{g-Q}$; 
and, therefore, $\tilde{H}_{j-A}$ can be expected to 
make only relatively unimportant contributions to low-energy processes. This fact supports our choice of 
gauge-invariant quark and gluon fields as appropriate for formulating a 
gauge-invariant description of QCD dynamics.\bigskip

We will also make some clarifying remarks about the relation between two representations used in this and 
earlier work:\cite{CBH2,BCH3} 
the ${\cal C}$ and the ${\cal N}$~representations. As we pointed out in earlier work,\cite{CBH2,BCH3} the 
Gauss's law operator ${\hat {\cal G}}^{a}({\bf r})=\partial_{i}\Pi_{i}^{a}({\bf r})+gf^{abc}A_{i}^{b}({\bf r})
\Pi_{i}^{c}({\bf r})+j_{0}^{a}({\bf r})$ with $j^a_{0}({\bf{r}})=g\,\,\psi^\dagger({\bf{r}})\,
{\textstyle\frac{\lambda^a}{2}}\,\psi({\bf{r}})\;,$ and the ``pure glue'' Gauss's law operator 
${\cal G}^{a}({\bf r})=\partial_{i}\Pi_{i}^{a}({\bf r})+gf^{abc}A_{i}^{b}({\bf r})\Pi_{i}^{c}({\bf r})$ are 
unitarily equivalent, so that ${\cal G}^a$ may be taken to represent ${\hat {\cal G}}^{a}({\bf r})$ in a different, 
unitarily equivalent representation. We refer to the representation in which ${\hat {\cal G}}^{a}({\bf r})$ is the 
Gauss's law operator, the ${\cal C}$ representation; and the representation in which 
${\cal G}$ represents the {\em entire} Gauss's law 
operator, with the color-charge density $j^a_{0}({\bf{r}})$ included (though implicitly only) 
the ${\cal N}$~representation. The unitary equivalence is expressed as\cite{CBH2} 
\begin{equation}
{\hat {\cal G}}^{a}({\bf r})
={\cal{U}}_{\cal{C}}\,
{\cal G}^{a}({\bf r})\,{\cal{U}}^{-1}_{\cal{C}}\,.
\label{eq:Gtrans}
\end{equation} 
The Gauss's law operator is the generator of infinitesimal gauge transformations; and the criterion for the 
gauge invariance of an operator ${\xi}$ is that it commute with the Gauss's law operator. 
In this work, when we use the Gauss's law operator to determine gauge invariance, it is important to distinguish
between the Gauss's law operators in the ${\cal C}$ and the ${\cal N}$~representations. 
An operator ${\xi}$ represents two different quantities
in the two representations. The quantity represented by ${\xi}$ in the  ${\cal C}$ representation has 
the form ${\cal{U}}^{-1}_{\cal{C}}\,{\xi}\,{\cal{U}}_{\cal{C}}$ in the ${\cal N}$~representation. But 
${\xi}$, when appearing in the ${\cal N}$~representation, refers to a different quantity, whose form 
in the ${\cal C}$ representation would be ${\cal{U}}_{\cal{C}}\,{\xi}\,{\cal{U}}^{-1}_{\cal{C}}$.
Since the quark field $\psi$ trivially commutes with ${\cal G}^{a}({\bf r})$, 
 $\psi$ is manifestly gauge-invariant in the 
${\cal N}$~representation. The unitary operator ${\cal{U}}_{\cal{C}}$ transforms the quark field $\psi$ so that 
\begin{equation}
{\psi}_{\sf GI}({\bf{r}})={\cal{U}}_{\cal C}\,\psi({\bf{r}})\,{\cal{U}}^{-1}_{\cal C}=
V_{\cal{C}}({\bf{r}}){\psi}({\bf{r}})
\label{eq:psiGI}
\end{equation}
is the form that the gauge-invariant gauge field takes in the more usual ${\cal C}$ representation. The unitary 
transformation given in Eq. (\ref{eq:psiGI}) has no effect on the gauge field. But the transformation it effects 
on the quark field is significant, because it allows us to use the quark field ${\psi}({\bf{r}})$ to represent the 
gauge-invariant quark field in the ${\cal N}$~representation. The unitary operator 
$V_{\cal{C}}({\bf{r}})$ is a functional of gauge fields and of the Gell-Mann 
matrices $\lambda^h\,$; its structure was discussed extensively in 
Ref.\cite{CBH2}. Since, in the ${\cal N}$~representation, ${\psi}({\bf{r}})$ is the expression for that same 
gauge-invariant field that is described by ${\psi}_{\sf GI}({\bf{r}})$ in the ${\cal C}$ 
representation, $V_{\cal{C}}({\bf{r}})$ 
is implicitly included in the quark field ${\psi}$ in the ${\cal N}~$representation. 
In the ${\cal N}$~representation, ${\psi}({\bf{r}})$ therefore implicitly 
consists of glue as well as quark field components. When the quark field  is gauge-transformed
within the ${\cal C}$ representation, ${\psi}({\bf{r}}){\rightarrow}{\exp}[ig({\lambda}^h/2)
{\chi}^h({\bf{r}})]{\psi}({\bf{r}})$,
where ${\chi}^h({\bf{r}})$ is an arbitrary time-independent c-number field in the adjoint representation of SU(3); 
that same gauge transformation transforms $V_{\cal{C}}({\bf{r}})
{\rightarrow}V_{\cal{C}}({\bf{r}}){\exp}[-\,ig({\lambda}^h/2){\chi}^h({\bf{r}})]$, so 
that ${\psi}_{\sf GI}({\bf{r}})$
is gauge-invariant.\cite{mario} In the ${\cal N}$~representation, the gauge invariance of 
${\psi}({\bf{r}})$ is a trivial 
consequence of the structure of the Gauss's law operator. \bigskip
 
These considerations have important consequences for the behavior of charge and current densities under 
gauge transformations. As a simple illustrative example, we consider the electron field operator $\psi$  
in QED in the temporal gauge. Under a gauge transformation, 
$A_{\mu}{\rightarrow}A_{\mu}-\partial_{\mu}\chi$, and the corresponding change in $\psi$ is 
${\psi}{\rightarrow}{\exp}(ie{\chi}){\psi}$. There are a number of ways of constructing gauge-invariant electron field operators 
in QED,\cite{khelqed} but one that is very useful for this discussion was provided by Dirac,\cite{dirac} who
defined the gauge-invariant electron field ${\psi}_{\sf GI}={\exp}[-ie(1/{\partial}^2){\partial}_iA_i\,]\,{\psi}$.
Under a gauge transformation, compensating changes occur in $\psi$ and in $(1/{\partial}^2){\partial}_iA_i$, 
so that ${\psi}_{\sf GI}$ remains gauge-invariant. The electric charge density operator can be represented either as 
$e{\psi}^{\dagger}_{\sf GI}{\psi}_{\sf GI}$ or as $e{\psi}^{\dagger}{\psi}$. Since the longitudinal photons (they are 
zero-norm ghosts) used to dress the electron field to make it gauge-invariant are electrically neutral, they 
do not affect the electric charge density, and the two expressions, $e{\psi}^{\dagger}_{\sf GI}{\psi}_{\sf GI}$ and 
$e{\psi}^{\dagger}{\psi}$, are identical. \bigskip

To discuss the non-Abelian case, we will use the SU(2) version of QCD, 
because the difference between the Abelian and
the non-Abelian cases can be illustrated very graphically in that system.  
We therefore use the SU(2) version of $V_{\cal{C}}({\bf{r}})$ in Eq. (\ref{eq:psiGI}), which is given by 
$V_{\cal{C}}({\bf{r}})_{SU(2)}={\exp}[-ig({\vec \tau}/2){\cdot}{\vec Z}({\bf{r}})]$, where 
${\vec Z}({\bf{r}})={(1/\partial}^2){\partial}_i{\vec B}_i$, and where ${\vec B}_i$ is a complicated 
functional of gauge fields, all of which commute with each other and with the spinor field $\psi$.~\cite{{CBH2}}
We can identify the gauge-invariant color charge density in the ${\cal C}$ representation of this SU(2) model of QCD as  
$[{\vec j}_{0}]_{{\sf GI}}=g{\psi}^{\dagger}_{\sf GI}({\vec {\tau}}/2)\,{\psi}_{\sf GI}$. In this non-Abelian case, 
$[{\vec j}_{0}]_{{\sf GI}}$ is no longer identical to ${\vec j}_{0}$. When the substitution
${\psi}_{\sf GI}={\exp}[-ig({\vec \tau}/2){\cdot}{\vec Z}({\bf{r}})]\,\psi$ is made, the gauge-invariant 
$[{\vec j}_{0}]_{{\sf GI}}$ can be expressed in terms of ${\vec j}_{0}=g{\psi}^{\dagger}({\vec {\tau}}/2)\,{\psi}$ by  
\begin{equation}
[{\vec j}_{0}]_{{\sf GI}}=\left({\hat Z}{\cdot}{\vec j}_{0}\right){\hat Z}+{\hat Z}{\times}\left({\vec j}_{0}{\times}
{\hat Z}\right){\cos}(Z)+\left({\vec j}_{0}{\times}{\hat Z}\right){\sin}(Z)\,,
\label{eq:ymcharge}
\end{equation}
where ${\hat Z}$ is the unit vector ${\hat Z}({\bf{r}})={\vec Z}({\bf{r}})/Z({\bf{r}})$.
Eq. (\ref{eq:ymcharge}) shows that in this case, and in its SU(3) version, dressing the 
gauge-dependent quark field with the gluons required to make it
gauge-invariant, does affect the color charge density, since the gluons themselves carry color. Under the infinitesimal 
gauge transformation  ${\delta}{\vec A}_i={\partial}_i{\delta}{\vec\chi}+g{\vec A}_i{\times}{\delta}{\vec \chi}$, 
the change in ${\vec j}_{0}$ is exactly compensated by the change in ${\hat Z}$,\cite{CBH2,mario}
so that $[{\vec j}_{0}]_{{\sf GI}}$ 
remains untransformed, and maintains its orientation in the ---in this instance SU(2) --- color space. 
The fact that a quantity 
is gauge-invariant does not mean that it has no preferred orientation in color space; 
it does mean that, whatever orientation 
it has, will not be altered by a gauge transformation. The same remark applies to the gauge-invariant gauge field
$A_{{\sf GI}\,i}^{a}({\bf r})$ as well. \bigskip

In view of the relation between  $[{\vec j}_{0}]_{{\sf GI}}$ and ${\vec j}_{0}$, it is particularly important to 
realize that ${\vec j}_{0}$ is the form that $[{\vec j}_{0}]_{{\sf GI}}$ takes in the ${\cal N}$~representation.  
The color-charge density $j^a_{0}({\bf{r}})$ as well as the color current density 
$j^a_{i}({\bf{r}})$ therefore are gauge-invariant operators in the ${\cal N}$
representation, and implicitly include glue as well as quark ingredients 
to constitute gauge-invariant color-charge and color-current densities. It is important to emphasize this point. 
The quantity that is represented by $j^a_{0}({\bf{r}})$ in the ${\cal C}$ representation is different from the 
quantity that is represented by $j^a_{0}({\bf{r}})$ in the ${\cal N}$~representation. $j^a_{0}({\bf{r}})$, 
when it appears in the ${\cal C}$ representation, 
consists of quark field components only, and transforms gauge-covariantly 
({\em not} gauge-invariantly) under an infinitesimal gauge transformation, in the form 
\begin{equation}
\delta\left[{j^a_{0}({\bf{r}})}\right]_{{\cal C}-rep}=
gf^{abe}\left[j^b_{0}({\bf{r}})\right]_{{\cal C}-rep}\delta{\chi}^e\,.
\label{eq:Ctrans}
\end{equation}
But $j^a_{0}({\bf{r}})$, when it appears in the ${\cal N}$~representation, is gauge invariant! 
A gauge-invariant field tensor can also be defined 
as\footnote{In this non-relativistic notation, $V_i$ refers to a contravariant vector and $\partial_i$ refers to the 
covariant derivative. This convention is used extensively throughout this work.}
\begin{equation}
F_{{\sf GI}\,ij}^{a}({\bf r})=\partial_jA_{{\sf GI}\,i}^{a}({\bf r})-\partial_iA_{{\sf GI}\,j}^{a}({\bf r})-
gf^{abc}A_{{\sf GI}\,i}^{b}({\bf r})A_{{\sf GI}\,j}^{c}({\bf r})
\label{eq:FIJGI}
\end{equation}
and then $F_{{\sf GI}\,ij}^{a}({\bf r})\frac{{\lambda}^a}{2}=F_{ij}^{a}({\bf r})V_{\cal{C}}({\bf{r}})
\frac{{\lambda}^a}{2}V^{-\,1}_{\cal{C}}({\bf{r}})$
and $F_{{\sf GI}\,ij}^{a}({\bf r})F_{{\sf GI}\,ij}^{a}({\bf r})=F_{ij}^{a}({\bf r})F_{ij}^{a}({\bf r})\,.$
Similarly, ${\sf K}^d_g({\bf r})$ and $\Pi^{a}_{i}({\bf r})\Pi^{a}_{i}({\bf r})$ are 
gauge-invariant quantities, so that {\em all}~the terms
appearing in ${\tilde H}$ are gauge-invariant. In Ref.\cite{CBH2}, we 
verified that the expressions we obtained for the gauge-invariant fields --- gauge as well as quark fields --- 
by expanding  ${V}_{\cal C}$ to arbitrary orders, agree with the expressions arrived at perturbatively by 
Lavelle and McMullan.\cite{lavelle2,lavelle3}\bigskip

The fact that, in the ${\cal N}$~representation, 
 $j^a_{0}({\bf{r}})$ and $j^a_{i}({\bf{r}})$ represent gauge-invariant charge and current densities 
respectively, enables us to use them to represent physical observables.  This makes it very convenient to 
formulate this work in the ${\cal N}$~representation.
It would, in principle, have been possible to carry out the investigation we are presenting here in either the 
${\cal C}$ or the ${\cal N}$~representation. But the use of the ${\cal N}$~representation 
is a very powerful tool for expressing the Hamiltonian in this gauge-invariant formulation. 
We will therefore proceed with this discussion, 
by using the ${\cal N}$~representation to examine how the  nonlocal interactions $H_{Q-Q}$ and $H_{g-Q}$ 
could provide a mechanism for the 
confinement of quarks when these are not in color-singlet configurations. A number of 
authors have discussed the importance of implementing Gauss's law in the quantization of QCD and 
Yang-Mills theory.\cite{goldjack,jackiw,khymtemp,rossinp,lands,karanair} 
And other authors have also speculated
that Gauss's law might be responsible for color confinement in QCD, and that, therefore,  gauge-invariant 
degrees of freedom might be necessary to make color confinement 
manifest.\cite{johnson4,lavelle2,lavelle3,johnson2,johnson3,lavelle5,stoll,lenz1,lenz2} 
A somewhat similar expression to our Eq. (\ref{eq:HQCDNVCRL}) --- later expressed as Eq. (\ref{eq:HQQ}) ---
has been given by 
T. D. Lee\cite{lee} and by Christ and Lee,\cite{christlee} but not with the same gauge-invariant gauge field 
$A^{\delta}_{\sf GI}$ between the inverse Laplacians that appears in our work. The gauge-invariant gauge 
field $A^{\delta}_{\sf GI}$, positioned between the inverse Laplacians,  is instrumental in providing for 
gauge invariance of the interaction Hamiltonian. In Ref.\cite{christlee}, unitary operators ($u$ and ${\cal U}$) 
transform the spinor (quark) fields so as to bring them into compliance with Gauss's law in a formulation in which 
Gauss's law is assumed to already be implemented in the gauge sector. Our work, reported in Ref.\cite{CBH2}, is 
based on the initial implementation of Gauss's law for the gauge sector, so that the functionals we construct,
that correspond to the angle variables in $u$ and ${\cal U}$ in Ref.\cite{christlee}, 
are explicitly given in terms of operator-valued gauge fields. These 
explicit expressions, which we obtained by  implementing Gauss's law for the gauge sector, are required to establish 
gauge invariance of the quark and gluon fields and of the color charge and current densities.
In spite of the important differences between the non-Abelian analogs to the Coulomb 
interaction that appear in Refs.\cite{lee,christlee}
and the one given in our Eqs. (\ref{eq:HQCDNVCRL}), (\ref{eq:HQQ})and (\ref{eq:Lexp}),  
the fact that nonlocal interactions of the same general structure
appear in these two treatments, which are quite different in their objectives and in the technical 
procedures used, supports the idea that such nonlocal interactions can have an important role in QCD dynamics.  \bigskip

We do not give a proof, in this work, that the long-range interactions $H_{Q-Q}$ and $H_{g-Q}$ confine quarks. 
We do present arguments, however, that these
nonlocal interactions,  which arise naturally when the QCD Hamiltonian is represented in terms of 
gauge-invariant fields, 
are interesting candidates for describing low-energy QCD dynamics, and that these nonlocal forces might have 
an important role in the confinement of quarks and the non-confinement of 
color-singlet configuration of quarks. \bigskip

\section{Quark Confinement, Color Transparency, and the Nonlocal Interaction}
\label{sec:Conftrans}
We will here examine to what extent the structure of ${\cal K}_0^b({\bf r})$ can be responsible for 
the confinement of quarks, or of  wave packets composed of quarks. To facilitate this discussion, we 
will display ${\cal K}_0^b({\bf r})$ in the expanded form
\begin{eqnarray} 
{\cal K}_0^b({\bf r})&&=
-j^b_{0}({\bf r})+(-g)\,f^{{\delta}_{(1)}ba}{A_{{\sf GI}\,i}^{{\delta}_{(1)}}({\bf{r}})\partial_{i}
\int\frac{d{\bf x}}{4{\pi}|{\bf r}-{\bf x}|}\,j^a_{0}({{\bf x}}})+\nonumber \\
&&g^2\,f^{{\delta}_{(1)}bs_{(1)}}f^{s_{(1)}{\delta}_{(2)}a}{A}_{{\sf GI}\,i}^
{{\delta}_{(1)}}({\bf{r}})\partial_{i}
\int\frac{d{\bf y}}{4{\pi}|{\bf r}-{\bf y}|}\,{A}_{{\sf GI}\,j}^{{\delta}_{(2)}}({\bf y})
\partial_{j}\int\frac{d{\bf x}}{4{\pi}|{\bf y}-{\bf x}|}\,j^a_{0}({\bf x})+\cdots\nonumber \\
+&&(-g)^n\,f^{{\delta}_{(1)}bs_{(1)}}{\cdots}f^{s_{(n-2)}{\delta}_{(n-1)}s_{(n-1)}}
f^{s_{(n-1)}{\delta}_{(n)}a}{A}_{{\sf GI}\,i}^{{\delta}_{(1)}}({\bf{r}})\partial_{i}
\int\frac{d{\bf y}_{(1)}}{4{\pi}|{\bf r}-{\bf y_{(1)}}|}\,{\cdots}
\times\nonumber \\ 
&&A_{{\sf GI}\,{\ell}}^
{{\delta}_{(n-2)}}({\bf y}_{(n-3)})\partial_{\ell}
\int\frac{d{\bf y}_{(n-2)}}{4{\pi}|{\bf y}_{(n-3)}-{\bf y}_{(n-2)}|}\,{A}_{{\sf GI}\,j}^{{\delta}_{(n-1)}}
({\bf y}_{(n-2)}) \times \nonumber \\
&&\partial_{j}
\int\frac{d{\bf y}_{(n-1)}}{4{\pi}|{\bf y}_{(n-2)}-{\bf y}_{(n-1)}|}{A}_{{\sf GI}\,k}^{{\delta}_{(n)}}
({\bf y}_{(n-1)})\,\partial_{k}\int\frac{d{\bf x}}{4{\pi}|{\bf y}_{(n-1)}-{\bf x}|}\,
j^a_{0}({{\bf x}})+\cdots\;
\label{eq:calTexp}.
\end{eqnarray}
Eq.~(\ref{eq:calT}) can be substituted into Eq.~(\ref{eq:HQCDNVCRL}), and a 
partial integration  carried out, so that
$(1/2){\int}d{\bf r}{\textstyle\frac{\partial_{j}}
{\partial^{2}}}{\cal K}_0^b({\bf r}){\textstyle\frac{\partial_{j}}{\partial^{2}}}{\cal K}_0^b({\bf r})$ 
is transformed into 
$-(1/2){\int}d{\bf r}{\cal K}_0^b({\bf r})\left(1/{\partial^{2}}\right){\cal K}_0^b({\bf r})$.
\bigskip

The first question to consider is how the nonlocal interaction described in 
Eqs.~(\ref{eq:HQCDNVCRL}) and (\ref{eq:calTexp})
would depend on the color of the participating quark configurations. We 
turn our attention to Eq.~(\ref{eq:HQCDNVCRL}) and represent it as 
\begin{equation}
H_{Q-Q}=-\,\frac{1}{2}{\int}d{\bf r}{\cal K}_0^b({\bf r})\left(\frac{1}{\partial^{2}}\right)
{\cal K}_0^b({\bf r})=\frac{1}{2}{\int}d{\bf r}d{\bf x}\,{j}_0^b({\bf r})
{\cal F}^{ba}({\bf r},{\bf x}){j}_0^a({\bf x})\,.
\label{eq:HQQ}
\end{equation}
${\cal F}^{ba}({\bf r},{\bf x})$ is a Green function that is the non-Abelian analog of 
$1/\left(4{\pi}|{\bf r}-{\bf x}|\right)$, but which differs from its QED analog in that it does not only 
refer to spatial points, but also depends on the gauge-invariant gauge field $A_{{\sf GI}\,i}^{\delta}({\bf{r}})$. 
${\cal F}^{ba}({\bf r},{\bf x})$ can easily be read from  Eq.~(\ref{eq:calTexp}), and shown to be 
\begin{eqnarray} 
{\cal F}^{ba}({\bf r},{\bf x})&&=\frac{{\delta}_{ab}}{4{\pi}|{\bf r}-{\bf x}|}
+2g\,f^{{\delta}_{(1)}ba}\int\frac{d{\bf y}}{4{\pi}|{\bf r}-{\bf y}|}\,{A_{{\sf GI}\,i}^
{{\delta}_{(1)}}({\bf{y}})\,\partial_{i}\frac{1}{4{\pi}|{\bf y}-{\bf x}|}}\nonumber \\
-3g^2\,f^{{\delta}_{(1)}bs_{(1)}}&&f^{s_{(1)}{\delta}_{(2)}a}\int\frac{d{\bf y}_1}{4{\pi}|
{\bf r}-{\bf y}_1|}\,
{A}_{{\sf GI}\,i}^{{\delta}_{(1)}}({\bf{y}_1})\,\partial_{i}
\int\frac{d{\bf y}_2}{4{\pi}|{\bf y}_1-{\bf y}_2|}\,{A}_{{\sf GI}\,j}^{{\delta}_{(2)}}
({\bf y}_{2})\,\partial_{j}
\frac{1}{4{\pi}|{\bf y}_2-{\bf x}|}+\cdots\nonumber\\
(-1)^{(n-1)}(n+1)&&g^nf^{{\delta}_{(1)}bs_{(1)}}f^{s_{(1)}{\delta}_{(2)}s_{(2)}}{\cdots}f^{s_{(n-1)}{\delta}_{(n)}a}
\int\frac{d{\bf y}_1}{4{\pi}|{\bf r}-{\bf y}_1|}\,
{A}_{{\sf GI}\,i}^{{\delta}_{(1)}}({\bf{y}_1})\,\partial_{i}
\int\frac{d{\bf y}_2}{4{\pi}|{\bf y}_1-{\bf y}_2|}\,\times\nonumber \\
{A}_{{\sf GI}\,j}^{{\delta}_{(2)}}&&
({\bf y}_{2})\,\partial_{j}\int\frac{d{\bf y}_3}{4{\pi}|{\bf y}_2-{\bf y}_3|}\,
\cdots\int\frac{d{\bf y}_n}{4{\pi}|{\bf y}_{(n-1)}-{\bf y}_n|}\,
{A}_{{\sf GI}\,{\ell}}^{{\delta}_{(n)}}
({\bf y}_{n})\,\partial_{\ell}
\frac{1}{4{\pi}|{\bf y}_n-{\bf x}|}+\cdots\,.
\label{eq:Lexp}
\end{eqnarray}
${\cal F}^{ba}({\bf r},{\bf x})$ can be seen to be symmetric under the combined interchange ${\bf r}\rightleftharpoons{\bf x}$ 
and $a{\rightleftharpoons}b$, when we assume that ${A}_{{\sf GI}\,i}^{\delta}({\bf{y}})
{\rightarrow}0$ when $|{\bf{y}}|{\rightarrow}\infty$. Eqs. (\ref{eq:HQQ}) and (\ref{eq:Lexp}) define a 
nonlocal interaction between gauge-invariant quarks or packets of quarks that is an analog of the Coulomb interaction in QED.
${\cal F}^{ba}({\bf r},{\bf x})$ is determined not only by the inverse Laplacians, but also by the matrix elements of 
the gauge-invariant gauge fields ${A}_{{\sf GI}\,i}^{\delta}({\bf{y}})$ in the gluonic medium in which the quarks are 
immersed. Once the spatial dependence of ${\cal F}^{ba}({\bf r},{\bf x})$ in a particular state of quark-gluon 
matter is fixed,
it describes a force that acts on quark color charges at points ${\bf x}$ and ${\bf y}$ respectively. 
Particularly in cases in which quark-antiquark creation is supressed 
and in which the quarks are at rest, or nearly at rest --- as in the case of heavy, static quarks --- 
${\cal F}^{ba}({\bf r},{\bf x})$ can play a very similar role in QCD to the Coulomb interaction in QED.\bigskip

We will apply  Eq.~(\ref{eq:HQQ}) to a state consisting of two packets of quarks that are 
immersed in glue, but are well-separated from each other in the sense that the quarks in each of the two 
packets can be represented using 
a complete set of orbitals, and that the two sets of orbitals occupied by the quarks in the 
two packets have a negligible overlap. We will not attempt to 
make any quantitative models for how such separated packets arise, but will only explore 
the interactions of such separated packets once 
they have arisen. We will furthermore assume in this discussion that quark-antiquark pair 
creation and annihilation can be neglected or ``quenched''. 
 We consider the expectation value 
\begin{equation}
\langle qQ\,|\,H_{Q-Q}\,|\,Qq\rangle=\langle\,q_k\,\cdots\,q_1\,Q_n\,\cdots\,Q_1\,|\,H_{Q-Q}\,|\,
q_1\,\cdots\,q_k\,Q_1\,\cdots\,Q_n\,\rangle\,,
\label{eq:QQmatrix}
\end{equation} 
where $q_1\,\cdots\,q_k$ represents a set of quarks in orbitals $u_j({\bf r})$ and $Q_1\,\cdots\,Q_n$ a set 
of quarks in orbitals $U_i({\bf r})$, so that both sets of orbitals are localized, and 
the overlap of $u_j({\bf r})$ and $U_i({\bf r})$ is negligible. In 
$|\,q_1\,\cdots\,q_k\,Q_1\,\cdots\,Q_n\,\rangle$, a set of creation operators for quarks 
$q_1,\,\cdots\,q_k\,,\,Q_1,\,\cdots\,Q_n\,,$ are applied to a state $|g\rangle$, which represents a gluonic 
medium that obeys the Gauss's law (in the ${\cal N}$~representation)
\begin{equation} 
{\cal G}^a\,|g\rangle=0\,.
\label{eq:gstate}
\end{equation} 
In Ref.\cite{CBH2}, a procedure was developed for constructing states such as $|g\rangle$.  
Since, in the ${\cal N}$~representation, quark creation and annihilation operators commute with 
the Gauss's law operator ${\cal G}^a$, multiquark Fock states can be constructed by 
applying quark creation operators to $|g\rangle$ without invalidating Gauss's law applied to the 
resulting multiquark Fock state. \bigskip

Because quark and glue field operators commute, and because of the negligible overlap of of the quark orbitals $u_j({\bf r})$ and $U_i({\bf r})$, 
$\langle qQ\,|\,H_{Q-Q}\,|\,Qq\rangle$ simplifies, through cluster decomposition, to 
\begin{equation}
\langle qQ\,|\,H_{Q-Q}\,|\,Qq\rangle={\int}d{\bf r}d{\bf x}\,{\bar {\cal F}}^{ba}({\bf r},{\bf x}){\langle}q_{k}{\cdots}q_{1}|
j^b_{0}({\bf r})|q_{1}{\cdots}q_{k}{\rangle}
{\langle}Q_{n}{\cdots}Q_{1}|j^a_{0}({\bf x})|Q_{1}{\cdots}Q_{n}{\rangle}\,,
\label{eq:HQQclust}
\end{equation}
where ${\bar {\cal F}}^{ba}({\bf r},{\bf x})$ represents the expectation value 
${\bar {\cal F}}^{ba}({\bf r},{\bf x})={\langle}g|\,{\cal F}^{ba}({\bf r},{\bf x})\,|g\rangle$. 
\bigskip

In evaluating ${\int}d{\bf x}\,{\bar {\cal F}}^{ba}({\bf r},{\bf x}){j}_0^a({\bf x}),$ --- one of the integrals required 
for Eq.~(\ref{eq:HQQ}) --- 
 we can expand about a point ${\bf x}_0$ located where the orbitals $U_i({\bf r})$ are large in magnitude. The integral then can be 
expressed as the Taylor's series, in a kind of ``color multipole'' expansion,
\begin{equation} 
{\int}d{\bf x}\,\left\{{\bar {\cal F}}^{ba}({\bf r},{\bf x}_0)\,+X_i\partial_i
{\bar {\cal F}}^{ba}({\bf r},{\bf x}_0)+\frac{1}{2}\,X_iX_j\,\partial_i\partial_j
{\bar {\cal F}}^{ba}({\bf r},{\bf x}_0)\;+\;\;\cdots\right\}{j}_0^a({\bf x})
\label{eq:Lmultipole}
\end{equation}
where $X_i=(x-x_0)_i$ and $\partial_i=\partial /\partial x_i\,.$ When we perform the integration in 
Eq.~(\ref{eq:Lmultipole}), the first term contributes ${\bar {\cal F}}^{ba}({\bf r},{\bf x}_0)\,{\cal Q}^a\,,$ where 
${\cal Q}^a=\int\,d{\bf x}\,{j}_0^a({\bf x})$ --- the integrated ``color charge''. Since the color charge is the 
generator of infinitesimal rotations in SU(3) space, 
it will annihilate any multiquark state vector in a singlet color configuration. 
Multiquark packets in a singlet color configuration therefore are immune to the leading term of the nonlocal 
$H_{Q-Q}$. Color-singlet configurations of quarks are only subject to the color multipole terms, 
which act as color analogs to the Van der Waals interaction. 
\bigskip

The scenario that this model suggests is that the leading term in $H_{Q-Q}$, namely 
${\cal Q}^b{\bar {\cal F}}^{ba}({\bf r}_0,{\bf x}_0){\cal Q}^a$, for 
a quark color charge ${\cal Q}^a$ at ${\bf r}_0$ and 
another quark color charge ${\cal Q}^b$ at ${\bf x}_0$, as well as $H_{g-Q}$, which describes 
the coupling between quark color charge and color-bearing gluonic matter, 
are  responsible for the confinement of 
quarks and packets of quarks that are not in color-singlet configurations. In this scenario, 
the multipole terms, which are the only parts of $H_{Q-Q}$ that affect the dynamics of 
color-singlet quark configurations,  do not confine packets, but result in  
final-state interactions that act on color-singlets as these 
move through, or emerge from a gluonic medium.  To confirm this scenario, it would be 
necessary to evaluate ${\bar {\cal F}}^{ba}({\bf r},{\bf x})$ and its spatial derivatives. This requires 
knowledge of the spatial dependence of $A_{{\sf GI}\,i}^{\delta}({\bf r})$. In Refs.\cite{CBH2,BCH3}, 
relationships of
$A_{{\sf GI}\,i}^{\delta}({\bf r})$ to other gauge-invariant (and gauge-covariant) quantities 
were established that would, in principle, allow specification of $A_{{\sf GI}\,i}^{\delta}({\bf r})$  
(or, rather, the specification of the set of possible values of $A_{{\sf GI}\,i}^{\delta}({\bf r})$, 
since the nonlinearity of the equations that determine those relationships may well signify
that $A_{{\sf GI}\,i}^{\delta}({\bf r})$ does not have a single unique value). 
But the explicit construction of $A_{{\sf GI}\,i}^{\delta}({\bf r})$ is not within the scope of this current work. 
We will, rather, explore some features of the behavior of color-singlet quark packets
as these move through a gluonic medium, 
assuming that ${\bar {\cal F}}^{ba}({\bf r},{\bf x})$ is a reasonably well-behaved function of ${\bf r}$ and ${\bf x}$. 
 \bigskip

In order to examine the dynamics of color-singlet quark configurations as these move through a medium 
consisting of gauge-invariant glue and quark matter, 
we will explore  the effect of the multipole terms on ``small'' color-singlet packets. Small packets 
occupy a limited region of space in the interior of gluonic matter, so that 
${\bar {\cal F}}^{ba}({\bf r},{\bf x})$ does not vary significantly 
over the spatial domain in which the packet functions $u_j({\bf r})$ and $U_i({\bf x})$ make sizable contributions. 
Given the assumption that ${\bar {\cal F}}^{ba}({\bf r},{\bf x}_0)$ 
varies only gradually within a volume occupied by quark packets, the effect of these
color multipole forces on a packet of quarks in a color-singlet configuration becomes more significant as 
the packet increases in size. As small quark packets move through gluonic matter, they 
will experience only insignificant effects from the multipole contributions to $H_{Q-Q}$, 
since, as can be seen from 
Eq.~(\ref{eq:Lmultipole}), the factors $X_i\,$, $X_iX_j$, ${\cdots}$, $X_{i(1)}{\cdots}X_{i(n)}$,
keep the higher order multipole terms from making significant contributions to 
${\int}d{\bf x}\,{\bar {\cal F}}^{ba}({\bf r},{\bf x}){j}_0^a({\bf x})$ when they are integrated 
over small packets of quarks. 
As the size of the quark packets increases, the regions
over which the multipoles are integrated also increases, and the effect of the multipole interactions
on the color-singlet packets can become larger. 
This dependence on packet size of the final-state interactions experienced by color-singlet states ---
$i.\,e.$ the increasing importance of final-state interactions as color-singlet packets grow in size --- 
is in qualitative agreement with the characterizations of color transparency and color coherence 
given by Miller and by Jain, Pire and Ralston.~\cite{miller,ralston} 
Eqs.(\ref{eq:HQQ}), (\ref{eq:Lexp}) and (\ref{eq:Lmultipole}) generalize the multipole expansion of a 
time-independent 
nonlocal interaction from the electromagnetic case, in which the leading term vanishes 
for a neutral object and 
leaves a residue of higher order multipoles, to QCD, in which color neutrality --- an attribute of 
color-singlet states --- corresponds to electrical neutrality in electrodynamics.  In the 
language of Jain, Pire and Ralston,~\cite{ralston} our analysis provides a model for how color neutrality protects a 
quark packet from the color-monopole force, with the result that 
such color-singlet packets can survive --- when the region over which the multipole interaction is integrated 
is small enough and the effect of ${\bar {\cal F}}^{ba}({\bf r},{\bf x})$ on it coherent enough ---  
to become asymptotic states observed in studies of exclusive processes. \bigskip

As previously mentioned, quark and gluon color is also coupled directly, in the form 
\begin{equation}
H_{g-Q}=-\int d{\bf r}d{\bf x}\left[{\sf K}_g^b({\bf r})\left(4{\pi}|({\bf r}-{\bf x})|\right)^{-1}
{\cal K}_0^b({\bf x})
+{\cal K}_0^b({\bf x})\left(4{\pi}|({\bf r}-{\bf x})|\right)^{-1}{\sf K}_g^b({\bf r})\right]\,. 
\label{eq:gqcouple}
\end{equation}
Whether our observations about the behavior of quark packets coupled to each other by $H_{Q-Q}$ 
also apply to quark packets coupled to gluonic matter directly by $H_{g-Q}$, depends on the distribution of 
glue --- whether, when the integration in Eq.~(\ref{eq:gqcouple}) ranges over regions in which
${\bf x}{\approx}{\bf r}$, the expansion in Eq.~(\ref{eq:Lmultipole}) is valid. We have to defer more detailed 
discussion of this question until the dynamical equations that determine the distribution of glue in this 
gauge-invariant formulation of the theory have been solved. \bigskip

Another point of interest about Eq. (\ref{eq:HQQclust}) is the fact that the spatial dependence of the 
long-range nonlocal interaction described by $H_{Q-Q}$  
 is largely determined by ${\bar {\cal F}}^{ba}({\bf r},{\bf x})$, which represents the expectation value 
${\bar {\cal F}}^{ba}({\bf r},{\bf x})=
{\langle}g|{\cal F}^{ba}({\bf r},{\bf x})|g\rangle$. And $|g\rangle$ 
implements the Gauss's law given in Eq. (\ref{eq:gstate}), in which the Gauss's law operator 
includes only the gluon color charge density. We observe, therefore, that the gluonic medium has a far more
important role in determining the long-range behavior of the nonlocal interaction $H_{Q-Q}$ than the quarks, 
which are acted on by $H_{Q-Q}$ (and by $H_{g-Q}$), but which have no role in transmitting these interactions. 
This result can therefore connect our work to other models for understanding  
color confinement. If it turns out that quarks, and not only gluons, are essential in transmitting confining forces 
from one group of quarks to another, then $H_{Q-Q}$ and $H_{g-Q}$ cannot be the only mechanisms for 
quark confinement. However, if other methods of analysis corroborate that the gluonic medium has the primary role
in effecting quark confinement, than $H_{Q-Q}$ and $H_{g-Q}$ become more interesting candidates as descriptions
of confining forces.

\section{non-perturbative approach to the nonlocal interaction in an SU(2) model of QCD}
\label{sec:nonpert}
In order to pursue the analysis undertaken in Section \ref{sec:Conftrans}, and to demonstrate a procedure for evaluating $H_{Q-Q}$ 
nonperturbatively, we will make use  
of Yang-Mills theory --- the SU(2) version of this model --- for which the structure constants 
$f^{{\delta}ba}$ are ${\epsilon}^{{\delta}ba}$. 
We will also ignore correlations among gauge-invariant gauge fields, and replace all  $A_{{\sf GI}\,i}^{\delta}({\bf{r}})$
in Eq. (\ref{eq:Lexp}) with the corresponding ${\langle}A_{{\sf GI}\,i}^{\delta}({\bf{r}})\rangle=
{\langle}g|\,A_{{\sf GI}\,i}^{\delta}({\bf{r}})\,|g\rangle$.
Furthermore, we will model ${\langle}A_{{\sf GI}\,i}^{\delta}({\bf{r}})\rangle$, 
which are transverse fields in the adjoint 
representation of SU(2), as the manifestly transverse ``hedgehog'' configuration
\begin{equation}
{\langle}A_{{\sf GI}\,i}^{\delta}({\bf{r}})\rangle={\epsilon}^{ij{\delta}}r_j{\phi}(r).
\label{eq:GISU2}
\end{equation}
Although there is no reason to believe that the {\em ansatz}~given in Eq.~(\ref{eq:GISU2}) follows 
from the dynamical equations that determine $A_{{\sf GI}\,i}^{\delta}({\bf{r}})$, it is a convenient choice 
for examining to what extent the structure of ${\bar {\cal F}}^{ba}({\bf r},{\bf x})$ --- 
as distinct from the precise form of $A_{{\sf GI}\,i}^{\delta}({\bf{r}})$ ---                                  
enables us to nonperturbatively evaluate the infinite series given in Eq.~(\ref{eq:Lexp}). 
This simplified SU(2) model can also help us to identify the features 
of the gauge-invariant gauge field that might be 
significant for the confinement of quarks or color-bearing quark packets. \bigskip

When we substitute ${\epsilon}^{\vec{\alpha}\beta\gamma}_{(\eta)}$ for $f^{\vec{\alpha}\beta\gamma}_{(\eta)}$,  
and use Eq.~(\ref{eq:GISU2}) in Eq.~(\ref{eq:calTexp}), we find that the integral in Eq.~(\ref{eq:calTexp}) 
that is linear in the gauge-invariant gauge field, can be expressed as
\begin{equation}
(-g)\,{\epsilon}^{{\delta}_{(1)}ba}\langle{A_{{\sf GI}\,i}^{{\delta}_{(1)}}({\bf{r}})\rangle\partial_{i}
\int\frac{d{\bf x}_1}{4{\pi}|{\bf r}-{\bf x}_1|}\,j^a_{0}({{\bf x}_1}})=ig\,{\epsilon}^{{\delta}_{(1)}ba}{\phi}(r)
L^{{\delta}_{(1)}}{\int\frac{d{\bf x}_1}{4{\pi}|{\bf r}-{\bf x}_1|}\,j^a_{0}({{\bf x}_1}})
\label{eq:first},
\end{equation}
where $L^{\delta}$ represents the ${\delta}$-component of the orbital angular momentum.  
In the case of the term quadratic in the gauge-invariant gauge field,  we observe that 
\begin{eqnarray}
g^2&&\,{\epsilon}^{{\delta}_{(1)}bs_{(1)}}{\epsilon}^{s_{(1)}{\delta}_{(2)}a}
\langle{A}_{{\sf GI}\,i}^{{\delta}_{(1)}}({\bf{r}})\rangle\partial_{i}
\int\frac{d{\bf x}_2}{4{\pi}|{\bf r}-{\bf x}_2|}\,\langle{A}_{{\sf GI}\,j}^{{\delta}_{(2)}}
({\bf x}_{2})\rangle\partial_{j}\int\frac{d{\bf x}_1}{4{\pi}|{\bf x}_2-{\bf x}_1|}\,j^a_{0}({{\bf x}_1})=\nonumber\\
-g^2&&\,{\epsilon}^{{\delta}_{(1)}bs_1}{\epsilon}^{s_1{\delta}_{(2)}a}{\phi}(r)L^{{\delta}_{(1)}}
\int\frac{d{\bf x}_2}{4{\pi}
|{\bf r}-{\bf x}_2|}\,{\phi}(x_2)L^{{\delta}_{(2)}}\int\frac{d{\bf x}_1}{4{\pi}|{\bf x}_2-{\bf x}_1|}
\,j^a_{0}({{\bf x}_1}).
\label{eq:second}
\end{eqnarray}
$L^{{\delta}_{(2)}}$ can be shifted to the left of 
the ${\bf x}_2$ integration, by noting that for any reasonably well-behaved ${\psi}({\bf x}_2)$,
\begin{eqnarray}
\int&&\frac{d{\bf x}_2}{4{\pi}|{\bf r}-{\bf x}_2|}\,{\phi}(x_2){\epsilon}^{{\delta}_{(2)}ji} (x_{2})_i\,
\frac{\partial}{{\partial}(x_{2})_j}{\psi}({\bf x}_2)=-{\epsilon}^{{\delta}_{(2)}ji}{\int}d{\bf x}_2\,{\phi}(x_2)
\left( (x_{2})_i\,\frac{\partial}{{\partial}(x_{2})_j}\,
\frac{1}{4{\pi}|{\bf r}-{\bf x}_2|}\right){\psi}({\bf x}_2)\nonumber \\
=&&-{\epsilon}^{{\delta}_{(2)}ji}r_j\frac{\partial}{{\partial}r_i}{\int}d{\bf x}_2\,{\phi}(x_2)
\left(\frac{1}{4{\pi}|{\bf r}-{\bf x}_2|}\right){\psi}({\bf x}_2)=
-iL^{\delta_2}{\int}d{\bf x}_2\,{\phi}(x_2)
\left(\frac{1}{4{\pi}|{\bf r}-{\bf x}_2|}\right){\psi}({\bf x}_2),
\label{eq:ShiftL}
\end{eqnarray}
and that the identical procedure can be carried out on every term in the series given in Eq.~(\ref{eq:calTexp}), 
to yield an expression in which all orbital angular momentum operators are on the extreme left-hand-side of 
the expression, and are functions of $r_{\ell}$ and ${\partial}/{\partial}r_{{\ell}^{\prime}}$. We thus obtain 
\begin{equation}
{\cal K}_0^b({\bf r})=\sum_{n=0}^{\infty}i^n{\epsilon}^{{\vec \delta}ba}_{(n)}L^{\vec \delta}_{(n)}
{\Phi}^{a}_{(n)}({{\bf r}})=-j^b_{0}({\bf r})+\sum_{n=1}^{\infty}i^n
{\epsilon}^{{\vec \delta}ba}_{(n)}L^{\vec \delta}_{(n)}
{\Phi}^{a}_{(n)}({{\bf r}})
\label{eq:calTSU2}
\end{equation} 
where $L^{\vec \delta}_{(n)}=\prod_{i=1}^nL^{\delta_i}$ and ${\epsilon}^{{\vec \delta}ba}_{(n)}$ is given by 
the SU(2) version of Eq.~(\ref{eq:fproductN}) (with the convention that ${\epsilon}^{{\vec \delta}ba}_{(1)}=
{\epsilon}^{{\delta}ba}$,  ${\epsilon}^{{\vec \delta}ba}_{(0)}=-{\delta}_{ba}$, and $L^{\vec \delta}_{(0)}=1$),  
and 
\begin{eqnarray}
{\Phi}^{a}_{(n)}({{\bf r}})={\phi}(r)&&\int\frac{d{\bf y}_{(1)}}{4{\pi}|{\bf r}-{\bf y}_{(1)}|}{\phi}(y_1)
\int\frac{d{\bf y}_{(2)}}{4{\pi}|{\bf y}_{(1)}-{\bf y}_{(2)}|}{\phi}(y_2){\cdots}
\int\frac{d{\bf y}_{(\ell)}}{4{\pi}|{\bf y}_{(\ell-1)}-{\bf y}_{(\ell)}|}{\phi}(y_\ell){\cdots}\times\nonumber \\
&&\int\frac{d{\bf y}_{(n-1)}}{4{\pi}
|{\bf y}_{(n-2)}-{\bf y}_{(n-1)}|}{\phi}(y_{(n-1)})\int\frac{d{\bf y}_{(n)}}
{4{\pi}|{\bf y}_{(n-1)}-{\bf y}_{(n)}|}\,j^a_{0}({\bf y}_{(n)})\,,
\label{eq:Phi}
\end{eqnarray}
with the convention that ${\Phi}^{a}_{(0)}({{\bf r}})=j^a_{0}({{\bf r}})$.

Because of the simplicity of the SU(2) structure constants --- the Kronecker delta is the only symmetric 
structure constant in SU(2) --- it is possible to significantly simplify $i^n{\epsilon}^{{\vec \delta}ba}_{(n)}
L^{\vec \delta}_{(n)}$. It was previously pointed out that the SU(2) chain of structure constants 
$\epsilon^{\vec{\alpha}\beta\gamma}_{(n)}$ can be represented as
\begin{equation}
\epsilon^{\vec{\alpha}\beta\gamma}_{(n)}=(-1)^{\frac{n}{2}-1}
\delta_{\alpha[1]\alpha[2]}\,\delta_{\alpha[3]\alpha[4]}\,
\cdots\,\delta_{\alpha[n-3]\alpha[n-2]}\,
\epsilon^{\alpha[n-1]\beta b}\,
\epsilon^{b\alpha[n]\gamma}\;
\label{eq:fproductN2}
\end{equation}
and
\begin{equation}
\epsilon^{\vec{\alpha}\beta\gamma}_{(n)}=(-1)^{\frac{n-1}{2}}
\delta_{\alpha[1]\alpha[2]}\,\delta_{\alpha[3]\alpha[4]}\,
\cdots\,\delta_{\alpha[n-2]\alpha[n-1]}\,
\epsilon^{\alpha[n]\beta \gamma}\;
\label{eq:fproductN3}
\end{equation} 
for even and odd $n$ respectively.\cite{CBH2} It is a trivial consequence of 
Eqs.~(\ref{eq:fproductN2}) and (\ref{eq:fproductN3}) that $i^n{\epsilon}^{{\vec \delta}ba}_{(n)}
L^{\vec \delta}_{(n)}$ can be represented as 
\begin{equation}
i^n{\epsilon}^{{\vec \delta}ba}_{(n)}L^{\vec \delta}_{(n)}={\cal L}_{(n)}^{ba}=
{\sf A}_n\left(L^aL^b+L^bL^a\right)+{\sf B}_n{\delta}_{ab}+i{\sf C}_n{\epsilon}_{abj}L^j\,, 
\label{eq:ABC}
\end{equation} 
where ${\sf A}_n$, ${\sf B}_n$, and ${\sf C}_n$ only depend on numerical constants 
and on the Casimir operator $L^2$. From 
simple recursion relations, $i.\,e.$ ${\sf D}_n+{\sf D}_{(n-1)}={\sf D}_{(n-2)}L^2$ 
where ${\sf D}_n={\sf A}_n+{\sf C}_n=
{\sf B}_{(n+1)}/L^2$,
and ${\sf F}_n+{\sf F}_{(n+1)}+{\sf D}_n+{\sf D}_{(n-1)}L^2=0$ where ${\sf F}_n=
{\sf A}_n-{\sf C}_n\,$, we can obtain 
the following explicit expressions for ${\sf A}_n$, ${\sf B}_n$, and ${\sf C}_n$. 
\begin{equation}
{\sf A}_n=\frac{(-1)^n}{2}\sum_{s=0}^{\left[\frac{n-2}{2}\right]}\frac{(n-s-1)!}{(n-2s-2)!(s+1)!}L^{2s}
\;\;\;\mbox{for}\;\;\;n{\geq}2\,,
\label{eq:A}
\end{equation}
\begin{equation}
{\sf B}_n=(-1)^{(n-1)}\sum_{s=0}^{\left[\frac{n-2}{2}\right]}\frac{(n-s-2)!}{(n-2s-2)!(s)!}L^{2(s+1)}
\;\;\;\mbox{for}\;\;\;n{\geq}2\,,
\label{eq:B}
\end{equation}
and 
\begin{equation}
{\sf C}_n=\frac{(-1)^{(n-1)}}{2}\sum_{s=0}^{\left[\frac{n-1}{2}\right]}\frac{(n-s-1)!(n-4s-3)}
{(n-2s-1)!(s+1)!}L^{2s}\;\;\;\mbox{for}\;\;\;n{\geq}2\,,
\label{eq:C}
\end{equation}
where $[{(n-2)}/{2}]={(n-2)}/{2}$ when $n$ is even, and $[{(n-2)}/{2}]={(n-3)}/{2}$ when $n$ is odd. 
for $n<2$, ${\sf A}_0={\sf A}_1={\sf B}_1={\sf C}_0=0$, and ${\sf B}_0={\sf C}_1=-1$.  
Because the algebra of the elements 
$(L^aL^b+L^bL^a)$, ${\delta}_{ab}$ and $i{\epsilon}_{abj}L^j$ is closed under multiplication, 
it is easily shown --- using the notation 
introduced in Eq.~(\ref{eq:ABC}) --- that 
\begin{equation}
{\cal L}_{(n+k)}^{ba}=-{\cal L}_{(n)}^{bs}{\cal L}_{(k)}^{sa}
\label{eq:Lscriptalg}
\end{equation}
where the repeated index $s$ is summed, and where
\begin{eqnarray}
&&{\sf A}_{n+k}=\left(2L^2-\frac{5}{2}\right){\sf A}_{n}{\sf A}_{k}+\left({\sf A}_{n}
{\sf B}_{k}+{\sf B}_{n}{\sf A}_{k}\right)-
\frac{3}{2}\left({\sf A}_{n}{\sf C}_{k}+{\sf C}_{n}{\sf A}_{k}\right)-\frac{1}{2}{\sf C}_{n}{\sf C}_{k}\,,\nonumber \\
&&{\sf B}_{n+k}={\sf A}_{n}{\sf A}_{k}\,L^2+\left({\sf A}_{n}{\sf C}_{k}+
{\sf C}_{n}{\sf A}_{k}\right)L^2+{\sf B}_{n}{\sf B}_{k}+
{\sf C}_{n}{\sf C}_{k}\,L^2\,, \nonumber \;\;\;\mbox{and}\\
&&{\sf C}_{n+k}=-\left(2L^2-\frac{3}{2}\right){\sf A}_{n}{\sf A}_{k}+
\left({\sf B}_{n}{\sf C}_{k}+{\sf C}_{n}{\sf B}_{k}\right)+
\frac{1}{2}\left({\sf A}_{n}{\sf C}_{k}+{\sf C}_{n}{\sf A}_{k}\right)-\frac{1}{2}{\sf C}_{n}{\sf C}_{k}\,.
\label{eq:ABCprod}
\end{eqnarray}
Since there are only three independent coefficients in any ${\cal L}_{(n)}^{ba}$ -- 
$viz.$ ${\sf A}_{n}$, ${\sf B}_{n}$ and 
${\sf C}_{n}$ --- the following equations must always have a unique solution for the constants 
${\sf x}_i$ for $i =1{\rightarrow}3$:
\begin{eqnarray}
-{\sf A}_0&&={\sf x}_1{\sf A}_1+{\sf x}_2{\sf A}_2+{\sf x}_3{\sf A}_3\,, \nonumber\\
-{\sf B}_0&&={\sf x}_1{\sf B}_1+{\sf x}_2{\sf B}_2+{\sf x}_3{\sf B}_3\,,  \;\;\;\mbox{and}\nonumber\\
-{\sf C}_0&&={\sf x}_1{\sf C}_1+{\sf x}_2{\sf C}_2+{\sf x}_3{\sf C}_3\,,
\label{eq:ABCsum}
\end{eqnarray}
so that the linear combination 
\begin{equation}
{\cal L}_{(0)}^{ba}+{\sf x}_1{\cal L}_{(1)}^{ba}+{\sf x}_2{\cal L}_{(2)}^{ba}+{\sf x}_3{\cal L}_{(3)}^{ba}=0\,.
\label{eq:Lsum}
\end{equation} 
Moreover, Eq.~(\ref{eq:Lscriptalg}) has the effect of generalizing the validity of Eqs.~(\ref{eq:ABCsum}) 
and (\ref{eq:Lsum}) so that they apply to 
coefficients ${\sf A}_{n+i}\,$, ${\sf B}_{n+i}$ and ${\sf C}_{n+i}$ for any fixed $n$ and $i =0{\rightarrow}3$.
The effect of these identities and of the explicit values of  the three ${\sf x}_1$ is that 
\begin{equation}
{\cal L}_{(n+3)}^{ba}+2{\cal L}_{(n+2)}^{ba}+\left(1-L^2\right){\cal L}_{(n+1)}^{ba}-L^2{\cal L}_{(n)}^{ba}=0\,
\label{eq:scriptLvanish}
\end{equation} 
for any $n$. The Casimir operator, $L^2$, is treated like any constant in this analysis. 
\bigskip

Eq.~(\ref{eq:scriptLvanish}) enables us to obtain a differential equation for  
${\cal K}_0^b({\bf r})$ whose features --- including its order --- 
reflect the SU(2) algebra as expressed in Eq.~(\ref{eq:ABC}). We observe that, for $n{\geq}1$, 
\begin{equation}
{\nabla}^2\frac{1}{{\phi}(r)}{\Phi}^b_{(n)}=-{\Phi}^b_{(n-1)}\,,
\label{eq:PHINN}
\end{equation}
which, with the use of  Eq.~(\ref{eq:Lscriptalg}), leads to:
\begin{equation}
{\nabla}^2\frac{1}{{\phi}(r)}\left[{\cal K}_0^b({\bf r})+j^b_{0}({\bf r})\right]=
{\cal L}_{(1)}^{bs}{\cal K}_0^s({\bf r}),
\label{eq:LaplaceK1}
\end{equation}
\begin{equation}
{\nabla}^2\frac{1}{{\phi}(r)}\left({\nabla}^2\frac{1}{{\phi}(r)}
\left[{\cal K}_0^b({\bf r})+j^b_{0}({\bf r})\right]\right)+
{\cal L}_{(1)}^{bs}{\nabla}^2\frac{1}{{\phi}(r)}j^s_{0}({\bf r})=
-{\cal L}_{(2)}^{bs}{\cal K}_0^s({\bf r})  \;\;\;\mbox{and} \;\;\;
\label{eq:LaplaceK2}
\end{equation}
\begin{eqnarray}
&&\left\{{\nabla}^2\frac{1}{{\phi}(r)}\left[{\nabla}^2\frac{1}{{\phi}(r)}\left({\nabla}^2\frac{1}{{\phi}(r)}
\left[{\cal K}_0^s({\bf r})+j^b_{0}({\bf r})\right]\right)\right]+
{\cal L}_{(1)}^{bs}{\nabla}^2\frac{1}{{\phi}(r)}\left({\nabla}^2
\frac{1}{{\phi}(r)}j^s_{0}({\bf r})\right)\right.\nonumber \\
&&\left.\;\;\;\;\;\;\;\;\;\;\;\;\;\;\;\;\;\;\;\;\;\;\;\;\;\;\;\;\;\;\;\;\;\;\;
\;\;\;\;\;\;\;\;\;\;\;\;\;\;\;\;\;\;\;\;\;
\;\;\;-{\cal L}_{(2)}^{bs}{\nabla}^2\frac{1}{{\phi}(r)}j^s_{0}({\bf r})\right\}=
{\cal L}_{(3)}^{bs}{\cal K}_0^s({\bf r}).
\label{eq:LaplaceK3}
\end{eqnarray}
We can use Eq.~(\ref{eq:scriptLvanish}) to combine Eqs.~(\ref{eq:LaplaceK1})-(\ref{eq:LaplaceK3}) in such a manner 
that the right-hand sides vanish, 
to generate a sixth-order equation for ${\cal K}_0^b({\bf r})$, given below:

\begin{eqnarray}
&&{\nabla}^2\frac{1}{{\phi}(r)}\left[{\nabla}^2\frac{1}{{\phi}(r)}\left({\nabla}^2\frac{1}{{\phi}(r)}
{\cal K}_0^b({\bf r})\right)\right]-2{\nabla}^2\frac{1}{{\phi}(r)}
\left({\nabla}^2\frac{1}{{\phi}(r)}{\cal K}_0^b({\bf r})\right)+L^2{\cal K}_0^b({\bf r})+\nonumber \\
&&+\left(1-L^2\right){\nabla}^2\frac{1}{{\phi}(r)}
{\cal K}_0^b({\bf r}) 
=\left[-\left(1-L^2\right){\delta}_{ba}+\left(2{\cal L}_{(1)}^{ba}+{\cal L}_{(2)}^{ba}\right)\right]
{\nabla}^2\frac{1}{{\phi}(r)}
j^a_{0}({\bf r})	+ \nonumber \\
&&+\left(2{\delta}_{ba}-{\cal L}_{(1)}^{ba}\right){\nabla}^2\frac{1}{{\phi}(r)}\left({\nabla}^2\frac{1}{{\phi}(r)}
j^a_{0}({\bf r})\right) 
-{\nabla}^2\frac{1}{{\phi}(r)}\left[{\nabla}^2\frac{1}{{\phi}(r)}\left({\nabla}^2\frac{1}{{\phi}(r)}
j^b_{0}({\bf r})\right)\right].
\label{eq:DiffeqK}
\end{eqnarray}
Although Eqs.~(\ref{eq:calTSU2}) and (\ref{eq:Phi}) represent ${\cal K}_0^b({\bf r})$ as an infinite series, which 
we might expect to have to evaluate perturbatively, Eq.~(\ref{eq:DiffeqK}) is a differential equation that 
${\cal K}_0^b({\bf r})$ obeys as a whole. The derivation of Eq.~(\ref{eq:DiffeqK}) 
therefore has enabled us to bypass the 
perturbative representation of ${\cal K}_0^b({\bf r})$.\bigskip 

We will make the further {\em ansatz}~that ${\phi}(r)$
is a complex constant, ${\kappa}$. Then, we can write Eq.~(\ref{eq:DiffeqK}) in the form 
\begin{equation}
\left(\frac{{\nabla}^6}{{\kappa}^{2}}-2\frac{{\nabla}^4}{\kappa}+(1-L^2){\nabla}^2+
{\kappa}L^2\right){\cal K}_0^b({\bf r})=s^b({\bf r})
\label{eq:DiffeqKsimp}
\end{equation}
where $s^b({\bf r})$ is the source term 
\begin{equation}
s^b({\bf r})=\left\{\left[-\left(1-L^2\right){\delta}_{ba}+\left(2{\cal L}_{(1)}^{ba}+
{\cal L}_{(2)}^{ba}\right)\right]{\nabla}^2+ 
\left(2{\delta}_{ba}-{\cal L}_{(1)}^{ba}\right)\frac{{\nabla}^4}{\kappa}
-\frac{{\nabla}^6}{{\kappa}^2}\right\}j^b_{0}({\bf r}).
\label{eq:source}
\end{equation}
In Eq.~(\ref{eq:source}), $s^b({\bf r})$ has been expanded in inverse powers of $\kappa$, 
and the leading term has been kept 
$\kappa$-independent. Eq.~(\ref{eq:DiffeqKsimp}) is a differential equation in which 
${\cal K}_0^b({\bf r})$ is related to a 
source. The spatial dependence of ${\cal K}_0^b({\bf r})$ can be determined by 
finding a Green function, $G({\bf r},{\bf x})$, 
for which 
\begin{equation}
\left(\frac{{\nabla}^6}{{\kappa}^{2}}-2\frac{{\nabla}^4}{\kappa}+(1-L^2){\nabla}^2+
{\kappa}L^2\right)G({\bf r},{\bf x})=
-{\delta}({\bf r}-{\bf x}).
\label{eq:green}
\end{equation}
Because the differential 
operator in Eq.~(\ref{eq:green}) is explicitly dependent on the orbital angular momentum $L^2$, 
it is desirable to expand $G({\bf r},{\bf x})$ in terms of partial waves.  We express $G({\bf r},{\bf x})$ as 
\begin{equation}
G({\bf r},{\bf x})=\sum_{{\ell},m}g_{\ell}(r,x)Y^{\ast}_{{\ell},m}({\theta}_{x},
{\phi}_x)Y_{{\ell},m}({\theta}_{r},{\phi}_r)
\label{eq:greenexp}
\end{equation}
and find that $g_{\ell}(r,x)$ obeys 
\begin{equation}
{\cal D}_rg_{\ell}(r,x)=-\frac{1}{r^2}{\delta}(r-x)
\label{eq:greenpartial}
\end{equation}
where ${\cal D}_r$ is the differential operator 
\begin{equation}
{\cal D}_r=\left[\frac{1}{{\kappa}^2}\left({\nabla}_r^2-\frac{{\ell}({\ell}+1)}{r^2}\right)^3-\frac{2}{\kappa}
\left({\nabla}_r^2-\frac{{\ell}({\ell}+1)}{r^2}\right)^2
+\left[1-{\ell}({\ell}+1)\right]\left({\nabla}_r^2-\frac{{\ell}({\ell}+1)}{r^2}\right) +
{\kappa}{\ell}\left({\ell}+1\right)\right].
\label{eq:diffopD}
\end{equation}
where ${\nabla}_r^2={\partial}_r^2+(2/r){\partial}_r$. We expand $g_{\ell}(r,x)$ as 
\begin{equation}
g_{\ell}(r,x)=\frac{2}{\pi}\int_0^{\infty}{\sf g}_{\ell}({\alpha})j_{\ell}({\alpha}r)
j_{\ell}({\alpha}x){\alpha}^2d{\alpha}
\label{eq:greenfb}
\end{equation}
and adjust ${\sf g}_{\ell}({\alpha})$ so that 
\begin{equation}
{\cal D}_rg_{\ell}(r,x)=-\frac{2}{\pi}\int_0^{\infty}j_{\ell}({\alpha}r)j_{\ell}({\alpha}x){\alpha}^2d{\alpha}=
-\frac{1}{r^2}{\delta}(r-x),
\label{eq:greenbessel}
\end{equation}
which leads to 
\begin{equation}
{\sf g}_{\ell}({\alpha})=\frac{1}{\kappa}\left[\left({\frac{\alpha^2}{\kappa}+1}\right)
\left(\frac{\alpha^4}{\kappa^2}+\frac{\alpha^2}{\kappa}
-{\ell}({\ell}+1)\right)\right]^{-1}.
\label{eq:sfg}
\end{equation}
Since ${\kappa}$ is a complex constant, we can parameterize it as $\kappa=|{\kappa}|{\exp}[i{\beta}]$ and, 
for ${\ell}>0$, represent 
$g_{\ell}(r,x)$ as 
\begin{eqnarray}
g_{\ell}(r,x)=\frac{{\kappa}^2}{\pi}\int^{\infty}_{-\infty}\frac{j_{\ell}({\alpha}r)
j_{\ell}({\alpha}x){\alpha}^2d{\alpha}}
{\left({\alpha}-i\sqrt{|{\kappa}|}e^{i{\beta}/2}\right)\left({\alpha}+
i\sqrt{|{\kappa}|}e^{i{\beta}/2}\right)} \times \nonumber \\
\left[\left({\alpha}-{\sf a}e^{i{\beta}/2}\right)\left({\alpha}+
{\sf a}e^{i{\beta}/2}\right)
\left({\alpha}-i{\sf b}e^{i{\beta}/2}\right)\left({\alpha}+
i{\sf b}e^{i{\beta}/2}\right)\right]^{-1}
\label{eq:greenpole}
\end{eqnarray}
 where 
\begin{equation}
{\sf a}=\sqrt{|{\kappa}|\left(\frac{\sqrt{4{\ell}({\ell}+1)+1}-1}{2}\right)\;}\;\;\;\;\;\mbox{and}\;\;\;\;
{\sf b}=\sqrt{|{\kappa}|\left(\frac{\sqrt{4{\ell}({\ell}+1)+1}+1}{2}\right)\;}\;\;\;.
\end{equation} 
For ${\ell}=0$, the corresponding expression is 
\begin{equation}
g_{0}(r,x)=\frac{{\kappa}^2}{{\pi}rx}\int^{\infty}_{-\infty}\frac{{\sin}({\alpha}r){\sin}({\alpha}x)d{\alpha}}
{{\alpha}^2\left({\alpha}-i\sqrt{|{\kappa}|}e^{i{\beta}/2}\right)^2
\left({\alpha}+i\sqrt{|{\kappa}|}e^{i{\beta}/2}\right)^2} 
\label{eq:zeropole}
\end{equation}
The integrals in Eqs.~(\ref{eq:greenpole}) and (\ref{eq:zeropole}) are simple to evaluate, but only some 
of the features of the resulting expressions are relevant to this discussion. For ${\ell}>0$, the dominant 
behavior of all the $g_{\ell}(r,x)$ as $r$ and $x$ increase in size is to decay exponentially in $r$ and $x$. 
For ${\ell}=0$ the poles at ${\alpha}={\pm}i\sqrt{|{\kappa}|}e^{i{\beta}/2}$
result in contributions that similarly decay exponentially in $r$ and $x$ unless $e^{i{\beta}/2}={\pm}i$;
but the doubly degenerate pole at $\alpha=0$ produces the contribution
\begin{equation}
[g_{0}(r,x)]_{({\alpha}=0)}=\frac{1}{2rx}\left((r+x)-|r-x|\right).
\label{eq:degpole}
\end{equation}

The discussion presented in this section leads us to make the following observations: The series 
representation, in Eq.~(\ref{eq:calTSU2}), of ${\cal K}_0^b({\bf r})$ --- a quantity that encapsulates 
the nonlocal interactions between quark color-charge densities $j_0^{b}({\bf r})$ with each 
other and with gluonic color-charge ---  
together with the SU(2) identities given in Eq.~(\ref{eq:scriptLvanish}), lead to a differential 
equation for ${\cal K}_0^b({\bf r})$ with source terms that are functionals of $j_0^{b}({\bf r})$. 
The derivation of this differential equation --- Eq.~(\ref{eq:DiffeqK}) --- enables us to 
eliminate the need for a term-by-term iterative expansion of ${\cal K}_0^b({\bf r})$. We have arrived at
Eq.~(\ref{eq:DiffeqK})  by replacing the  SU(3) structure constants 
that apply to QCD with their corresponding SU(2) equivalents, and 
we have simplified Eq.~(\ref{eq:DiffeqK}) to the form given in Eq.~(\ref{eq:DiffeqKsimp}) by imposing
an {\em ad hoc}~ansatz that fixes the functional dependence of 
the gauge-invariant gauge field 
on spatial variables and SU(2) indices as shown in 
Eq.~(\ref{eq:GISU2}). In spite of the special assumptions that apply to this nonperturbative 
evaluation of ${\cal K}_0^b({\bf r})$, it can serve as a useful model for a similar approach applicable to a 
more realistic treatment of QCD with SU(3) structure constants 
and with gauge-invariant gauge fields that reflect more of the the dynamics of this theory. \bigskip

The solutions we obtained for our simplified SU(2) model --- 
Eqs.~(\ref{eq:greenfb})-(\ref{eq:degpole}) --- indicate that our simplifying assumptions lead to a form for 
the gauge-invariant gauge field that does not make $H_{Q-Q}$ a confining nonlocal interaction. 
The fact that Eq.~(\ref{eq:DiffeqKsimp})
is a sixth-order equation might have led us to anticipate that ${\cal F}^{ba}({\bf r},{\bf x})$ would 
be a confining interaction. Even fourth-order equations can lead to Green functions with linear potentials 
that confine.\cite{pdm} But, in this present case, the SU(2) structure constants lead to the pole structure for the 
${\ell}$-th partial wave shown in Eq.~(\ref{eq:greenpole}), with $\sqrt{{\kappa}\,{\ell}({\ell}+1)}$ 
acting like a mass term in the differential equation that defines the Green function. Green functions 
that confine with linear potentials, or even with more rapidly rising ones,  
require higher order degenerate poles. Moreover, if the 
Green function's exponential decrease with distance is to be avoided, the degenerate poles must be 
on the real-${\alpha}$ axis; and, if oscillatory behavior of the Green function is also to be avoided, 
the degenerate poles must be at 
the origin in the $\alpha$ plane.  In our  model, only $g_{0}(r,x)$ has a degenerate pole at the origin, 
and the pole is not degenerate to a sufficiently high order to support confinement. We could expect that 
the richer algebra of the SU(3) structure constants would lead to a higher order 
differential equation, and that the SU(3) Green functions therefore could lead to confinement --- 
perhaps through a more degenerate pole structure. 
But in order to determine whether that is the case, it will be necessary to 
find an expression for the gauge-invariant gauge field that  adheres more closely to the dynamics of the theory,
and is not based on an {\em ad hoc}~ansatz. \bigskip

To illustrate this discussion with a specific example, we consider a hypothetical case in which 
both $L^2$ and $1-L^2$ in Eq.~(\ref{eq:DiffeqKsimp}) vanish --- an obvious impossibility, since 
the eigenvalues of $L^2$ are quantized to a set of possible eigenvalues that forbid this. Nevertheless, 
the consequences of this assumption serve a useful illustrative purpose. For this hypothetical case, we obtain 
the following expression for  ${\bar g}_0(r,x)$, the ${\ell=0}$ component of the Green function: 
\begin{equation}
{\bar g}_{0}(r,x)=\frac{{\kappa}^2}{{\pi}rx}\int^{\infty}_{-\infty}\frac{{\sin}({\alpha}r){\sin}({\alpha}x)d{\alpha}}
{{\alpha}^4\left({\alpha}^2+2{\kappa}\right)}\,,
\label{eq:zeropolehyp}
\end{equation}
which includes a contribution from the quadruply degenerate pole at $\alpha=0$,
\begin{equation}
[{\bar g}_{0}(r,x)]_{({\alpha}=0)}=-\frac{\kappa}{24rx}\left((r+x)^3-|r-x|^3\right).
\label{eq:degpolehyp}
\end{equation}
The spatial dependence of $[{\bar g}_{0}(r,x)]_{({\alpha}=0)}$ is consistent with a 
confining potential for color-bearing quark packets. Although the dynamics of the model we are 
exploring does not lead to Eq.~(\ref{eq:degpolehyp}), the result may nevertheless serve a useful illustrative purpose. 
The model 
we are investigating in this section, which includes SU(2) structure constants and the ``hedgehog'' representation
 of the gauge-invariant
gauge field, with the spatial function ${\phi}(r)$ represented as the constant ${\phi}(r)=\kappa$, is 
itself a toy model used to represent the less tractable theory that has SU(3) structure constants and 
a gauge-invariant gauge field obtained from the dynamical equations presented in Refs.\cite{CBH2,BCH3}. 
It is relevant to inquire what changes can occur in the behavior of ${\cal K}_0^b({\bf r})$, and therefore 
also of the coupling term for quark color-charge densities, ${\cal F}^{ba}({\bf r},{\bf x})$, when 
changes are made in the equations that determine the Green function $G({\bf r},{\bf x})$, which
might well be duplicated in the full SU(3) version of QCD.\bigskip

The model proposed in this section, consisting of SU(2) structure constants and the 
hedgehog {\em ansatz}~for the gauge-invariant 
gauge field, does not provide us with much information about $H_{g-Q}$ --- 
the direct coupling between quark and glue color. 
When the gauge-invariant gauge field appears as part of ${\cal F}^{ba}({\bf r},{\bf x})$, 
it is evaluated in expressions that 
do not contain the operator $\Pi^{c}_{i}({\bf r})$ conjugate to the gauge-dependent 
gauge field. Ignoring field correlations, 
and replacing $A_{{\sf GI}\,i}^{\delta}({\bf{r}})$ with ${\langle}A_{{\sf GI}\,i}^{\delta}({\bf{r}})\rangle$ in 
${\cal F}^{ba}({\bf r},{\bf x})$, therefore is a legitimate and useful approximation. 
The model is not, however, as applicable 
to the representation of ${\sf K}_g^d({\bf r})=gf^{d\sigma e}\,{\sf Tr}\left[V_{\cal{C}}^{-1}({\bf{r}})
{\textstyle\frac{\lambda^e}{2}}V_{\cal{C}}({\bf{r}}){\textstyle\frac{\lambda^b}{2}}\right]
A_{{\sf GI}\,i}^{\sigma}({\bf r})\Pi^{b}_{i}({\bf r})$, in which $A_{{\sf GI}\,i}^{\delta}({\bf{r}})$ 
and $\Pi^{b}_{i}({\bf r})$
appear together. Our model does not include an expression for $\Pi^{b}_{i}({\bf r})$ that would be consistent with the 
hedgehog ansatz, or even respect the necessary commutation rules between the 
gauge field and its canonical adjoint. The 
effect of $H_{g-Q}$ on color confinement therefore remains to be addressed until the dynamics of 
the gauge-invariant gauge field have been more fully explored.

\section{Discussion}
\label{sect:Disc}
In this work, we have analyzed the nonlocal interaction that results
when QCD is formulated in terms of gauge-invariant quark and gluon operator-valued fields. We have shown 
that this nonlocal interaction involves the quark color-charge density in a way that is roughly analogous 
to the role of the electric charge density in the Coulomb interaction in a gauge-invariant formulation of
QED;  but the functional form of this nonlocal interaction --- the 
QCD analog of $1/(4{\pi}|{\bf r}-{\bf x}|)$ in QED --- is ${\cal F}^{ba}({\bf r},{\bf x})$, which is 
a nonlocal functional that depends not only  on spatial variables, but also 
involves the gauge-invariant gauge field. ${\cal F}^{ba}({\bf r},{\bf x})$ is an 
infinite series, in which the $n$-th order term contains the gauge-invariant gauge field to the $n$-th
power. But the series has such a regular structure, that an explicit form for the $n$-th order term 
can easily be written, without requiring knowledge of the lower order terms in the series. A nonperturbative 
treatment of this nonlocal interaction term is therefore not nearly as inaccessible as would be the case  for 
S-matrix elements in perturbative QCD. \bigskip

One feature of the nonlocal interaction between quark color-charge densities
is that the monopole color charge is its leading term, and that higher order multipole 
interactions --- Van der Waals type forces --- succeed it in a series expansion.
A color-charge monopole force as the leading term for quarks or color-bearing ensembles of quarks, and 
the consequence of that idea --- that color singlets, the natural QCD analogs of electrically neutral 
ensembles of charges, would not feel that force --- is not new. This idea --- and the term ``color neutrality'' to  
designate it ---  have, for example, been suggested by Jain, 
Pire and Ralston.\cite{ralston} These authors also have suggested that 
there is a connection between color transparency and the lesser importance of higher order 
Van der Waals-like multipoles for small-sized color singlet configurations. What is new in
our work is that these features of a nonlocal QCD interaction are no longer a conjecture motivated by 
phenomenology only, but are inherent in the representation of the QCD Hamiltonian  
in terms of gauge-invariant quark and gluon fields.\bigskip

The gauge-invariant form of the QCD Hamiltonian contains the kinetic energy terms for the gauge-invariant 
quark and gluon fields, the nonlocal interaction discussed in this paper, and the additional interaction 
term ${\tilde H}_{j-A}=-g\int{\psi^\dagger}({\bf r})\alpha_{i}{\textstyle\frac{\lambda^a}{2}}
\psi({\bf r})A_{{\sf GI}\,i}^{a}({\bf r})d{\bf r}$, which is the QCD analog of the QED interaction 
term $-e\int{\psi^\dagger}({\bf r})\alpha_{i}\psi({\bf r})A_{{\sf GI}\,i}\,({\bf r})d{\bf r}$. 
In QED, this latter term
describes the interaction of the electron current with the gauge-invariant gauge field 
(in QED, $A_{{\sf GI}\,i}\,({\bf r})$ is just 
the transverse component of the gauge field). 
In QED, this interaction couples electrons to the two helicity modes of the photons --- the propagating, observable 
quantized modes of the transverse electromagnetic vector potential.  As is well known, 
the current appearing in this interaction has 
a $v/c$ dependence, that makes it relatively unimportant in the low-energy regime, in which the Coulomb interaction
is, by far, the most important interaction between charged particles. Since the 
corresponding term in the QCD Hamiltonian 
similarly involves a current density --- the transverse color-current density in this case --- 
it is reasonable to expect that its $v/c$ dependence will also make it much less important than the 
nonlocal interaction between quark color-charge densities in the low-energy regime. The interaction term 
$H_{g-Q}+H_{Q-Q}$ therefore is a very interesting candidate low-energy limit of of the QCD interaction --- 
a nonlocal interaction term between quark color-charge densities with each other and with gluonic color,
that, in analogy with the Coulomb interaction in QED, would describe the most 
important features of QCD in the low-energy regime. \bigskip

It is important to make the following distinction between different parts of this work: 
On the one hand, there are relations like 
Eqs.~(\ref{eq:calTexp})-(\ref{eq:Lexp}),  which are exact. 
These relations are useful to the extent to which we have 
identified serviceable gauge-invariant fields that lead to a gauge-invariant Hamiltonian 
in which the nonlocal interactions that result 
from the implementation of gauge invariance are dominant in describing static quarks. These 
relations are not dependent on any approximations. They are direct and inevitable 
consequences of transforming the QCD Hamiltonian to a representation in which it is 
expressed in terms of the gauge-invariant
fields we have constructed. On the other hand, there are the contents of Section~\ref{sec:nonpert},
which serve an important illustrative purpose, but are 
dependent on a number of simplifying assumptions. Our discussion, in Section~\ref{sec:Conftrans}, of the 
implications of the nonlocal interaction for color confinement 
and for color transparency, are not dependent on the substitution of the SU(2) 
for the SU(3) algebra, or the {\em ansatz}~that the gauge-invariant gauge fields are uncorrelated or that they
have a hedgehog configuration, which we made in Section~\ref{sec:nonpert}. This part of the discussion 
in Section~\ref{sec:Conftrans} is dependent only on assumptions about the 
functional dependence of ${\bar {\cal F}}^{ba}({\bf r},{\bf x})$ on the spatial 
variables ${\bf r}$ and ${\bf x}$ , and about the sizes of the quark 
packets coupled by this nonlocal interaction. We have developed this model 
far enough so that we can explore the consequences of our model, 
given these assumptions. We cannot, at this point, establish that 
these assumed quark configurations will necessarily arise in the low-energy regime. \bigskip

The differential equation, Eq. (\ref{eq:DiffeqK}) and its special form Eq.(\ref{eq:DiffeqKsimp}), and the Green 
function solution of Eq. (\ref{eq:DiffeqKsimp})  given in Section~\ref{sec:nonpert}, depend on two 
simplifying assumptions --- the SU(2) algebra 
and the hedgehog configuration of the gauge-invariant gauge field. These results are 
important for establishing a pattern for nonperturbative treatments of the nonlocal 
interaction in QCD and for demonstrating the role of the 
SU(N) algebra in generating higher order differential equations for the Green 
function that connects quarks to each other. But the specific 
solutions --- Eqs.~(\ref{eq:greenpole})-(\ref{eq:degpole}) --- reported in Section III only apply to the ``toy model''
based on SU(2) structure constants and the hedgehog ansatz for the gauge-invariant gauge field.

\acknowledgements
This research was supported by the Department of Energy
under Grant No. DE-FG02-92ER40716.00.

\end{document}